\documentclass[prl,aps,twocolumn,preprintnumbers,showpacs,showkeys,superscriptaddress]{revtex4}
\def\bfl{\begin{flushleft}}
\def\efl{\end{flushleft}}
\def\bfr{\begin{flushright}}
\def\efr{\end{flushright}}
\def\bc{\begin{center}}
\def\ec{\end{center}}
\def\be{\begin{equation}}
\def\ee{\end{equation}}
\def\ba{\begin{eqnarray}}
\def\ea{\end{eqnarray}}
\def\baa#1{\begin{array}{#1}}
\def\eaa{\end{array}}
\def\bw{\begin{widetext}}
\def\ew{\end{widetext}}
\def\nn{\nonumber }

\def\text#1{\mbox{#1}}

\begin{document}

%\preprint{APS/123-QED}

\title{Conductivity rules in the Fermi and charge-spin separated liquid}

\author{Andrew Das Arulsamy}
%\cite{z}

\affiliation{ Condensed Matter Group, Division of Exotic Matter,
No. 22, Jalan Melur 14, Taman Melur, 68000 Ampang, Selangor DE,
Malaysia}

\date{\today}

%\cite{z}

%\date{1$^{st}$ January 2003}
%\date{~Received: 12 Jan 2001 [LANL] ~}
%\date{~Received: 26 May 2000 [PRL], 1 June 2000 [LANL] ~}
%\date{Received \today}

%\scriptsize%\footnotesize

\begin{abstract}
Ioffe-Larkin rule applies for the pure charge-spin separation
regardless of its dimensionality. Here, an extension to this rule
as a result of the coexistence of spinon, holon and electron as a
single entity in the 2-dimensional (2D) system is derived, which
is also in accordance with the original rule.
\end{abstract}


 \maketitle

%\narrowtext

%\small

\subsection{1. Introduction}

The coexistence of Fermi (electron) and charge (holon)-spin
(spinon) separated (F-CSS) liquid in 2D high-$T_c$ cuprate
superconductors have been reported in order to explain the $T$
dependence of magneto-thermo-electronic transport
properties~\cite{andrew1,andrew2}. Such coexistence strictly
requires the $ab$-plane and $c$-axis resistivity models
($\sigma^{-1}$) in the form of

\begin{eqnarray}
&&\sigma^{-1}_{ab} = \sigma^{-1}_s + \sigma^{-1}_h +
\sigma^{-1}_{s + h \to e} = \sigma^{-1}_s + \sigma^{-1}_h +
\gamma\sigma^{-1}_e \nn
\\&& \sigma^{-1}_c = \sigma^{-1}_e + \sigma^{-1}_{e
\rightleftharpoons s + h} = \sigma^{-1}_e + \beta[\sigma^{-1}_h +
\sigma^{-1}_s]. \label{eq:1}
\end{eqnarray}

The subscripts $e$, $s$ and $h$ represent the electrons, spinons
and holons respectively, while the subscripts $ab$ and $c$ denote
the $ab$-planes and $c$-axis respectively. $\gamma$ and $\beta$
however, are the experimental related constants of proportionality
(range from 0 $\to$ 1), which are associated with the contribution
of $c$-axis in $ab$-planes or vice versa~\cite{andrew1,andrew2}.
The term $\sigma^{-1}_{s + h \to e}$ is defined to be the
resistivity caused by the $s$ + $h$ $\to$ $e$ process occurring in
the $ab$-planes that gives rise to the electron-electron
scattering rate in $ab$-planes. If $s$ + $h$ $\to$ $e$ is
completely blocked in $ab$-planes then the electron's density in
the $ab$-planes is zilch and consequently, $\sigma^{-1}_{s + h \to
e}$ = 0. Any increment in $\sigma^{-1}_e$ also increases
$\sigma^{-1}_{s + h \to e}$. Therefore, $\sigma^{-1}_{s + h \to
e}$ $\propto$ $\sigma^{-1}_e$ (or $\sigma^{-1}_{s + h \to e}$ =
$\gamma\sigma^{-1}_e$).

In contrast, the term $\sigma^{-1}_{e \rightleftharpoons s + h}$
is defined to be the resistivity arises from the blockage in the
$e$ $\rightleftharpoons$ $s$ + $h$ processes. In other words,
$\sigma^{-1}_{e \rightleftharpoons s + h}$ is due to the
non-spontaneous $e$ $\to$ $s$ + $h$ and $s$ + $h$ $\to$ $e$
processes. These non-spontaneous processes imply that the spinons
and holons are not energetically favorable in $c$-axis while the
electrons are not energetically favorable in $ab$-planes. In
addition, any increment in $\sigma^{-1}_{ab}$ further increases
the magnitude of the blockage in $e$ $\rightleftharpoons$ $s$ +
$h$ processes that eventually leads to a larger $\sigma^{-1}_{e
\rightleftharpoons s + h}$. Consequently, $\sigma^{-1}_{e
\rightleftharpoons s + h}$ $\propto$ $\sigma^{-1}_{ab}$ (or
$\sigma^{-1}_{e \rightleftharpoons s + h}$ =
$\beta\sigma^{-1}_{ab}$). Simply put, the $e \rightleftharpoons s
+ h$ processes become increasingly difficult with increasing
$\sigma^{-1}_{ab}$. This proportionality can also be interpreted
as the additional scattering for the electrons to pass across
$ab$-planes. If $e$ $\rightleftharpoons$ $s$ + $h$ is spontaneous
then $\sigma^{-1}_{e \rightleftharpoons s + h}$ = 0.

\subsection{2. Theoretical details}

Here, Eq.~(\ref{eq:1}) is shown to be microscopically relevant
with the original Ioffe-Larkin's approach by using their effective
long range action ({\it S}) in the presence of electromagnetic
field ($A$), effective interactions between fermionic and bosonic
fields and gauge field ($a$). The mentioned action that describes
the Ioffe-Larkin formula is given by~\cite{ioffe4}

\begin{eqnarray}
S\{A,a\} &=& \frac{T}{2}\int
d\textbf{k}\sum_{\omega}\big\{[Q_sA(\omega,\textbf{k}) +
a(\omega,\textbf{k})] \nn
\\&& \times \Pi_s(\omega,\textbf{k})[Q_sA(\omega,\textbf{k}) + a(\omega,\textbf{k})]
\nn \\&& + [Q_hA(\omega,\textbf{k}) + a(\omega,\textbf{k})] \nn
\\&& \times \Pi_h(\omega,\textbf{k})[Q_hA(\omega,\textbf{k}) + a(\omega,\textbf{k})]\big\}. \label{eq:2}
\end{eqnarray}

The respective $\omega$ and $\textbf{k}$ represent the frequency
and the wave vector. $Q_{s}$ and $Q_{h}$ denote the arbitrary
charges of a spinon and a holon respectively, while $Q_{e}$ is the
charge of an electron. Note that Eq.~(\ref{eq:2}) ignores spinon
pairing and arbitrary charges have been assigned
accordingly~\cite{ichinose5}. Both spinons (fermions) and bosons
(holons) interact with $A$ and $a$. In this work, $S$ in
Eq.~(\ref{eq:2}) is rewritten by writing an additional interaction
generated by the electrons coexistence with spinons and holons,
which is explicitly given by

\begin{eqnarray}
\mathcal{S} = S\{A,a\} &+& \frac{T}{2}\int
d\textbf{k}\sum_{\omega}\big\{Q_eA(\omega,\textbf{k}) \nn
\\&& \times \Pi_{e \rightleftharpoons s + h}(\omega,\textbf{k})
Q_eA(\omega,\textbf{k})\big\}. \label{eq:3}
\end{eqnarray}

This additional term is zero to satisfy the principle of least
action and also to imply that the electrons are not a separate
entity in which, electrons flow is very much depends on spinons
and holons flow and vice versa. This single-entity requirement
will be discussed with appropriate limits shortly. The effective
Lagrangian that corresponds to $\mathcal{S}$ is actually given by

\begin{eqnarray}
&&\mathcal{L}[A,a] = \sum_{ij}a_i(\Pi^{ij}_s + \Pi^{ij}_h)a_j
\nonumber \\&& + 2\sum_{ij}a_i(Q_s\Pi^{ij}_s + Q_h\Pi^{ij}_h)A_j
\nonumber \\&& + \sum_{ij}A_i(Q_s^2\Pi^{ij}_s +
Q_h^2\Pi^{ij}_h)A_j \nonumber \\&& + \sum_{ij}
A_i\bigg[\frac{\big(\gamma +
\beta\big)Q_e^2\Pi^{ij}_s\Pi^{ij}_h\Pi^{ij}_e}{\gamma\Pi^{ij}_s\Pi^{ij}_h
+ \beta\Pi^{ij}_e\Pi^{ij}_s + \beta\Pi^{ij}_h\Pi^{ij}_e}\bigg]A_j
\nonumber
\\&& - \sum_{ij} A_i\bigg[
\frac{\big(\gamma +
\beta\big)Q_e^2\Pi^{ij}_s\Pi^{ij}_h\Pi^{ij}_e}{\gamma\Pi^{ij}_s\Pi^{ij}_h
+ \beta\Pi^{ij}_e\Pi^{ij}_s + \beta\Pi^{ij}_h\Pi^{ij}_e}\bigg]A_j.
\label{eq:4}
\end{eqnarray}

$\Pi_{s,h,e}$ denotes the response function for the spinons,
holons and electrons, respectively. The single entity scenario
allows electrons to pass across $ab$-planes with strong
interaction with spinon-holon flow. On the contrary, if the
electrons are an independent entity, not influenced by the spinons
and holons flow, then the action, $S_{ind}$ is simply given by

\begin{eqnarray}
S_{ind} = S\{A,a\} &+& \frac{T}{2}\int
d\textbf{k}\sum_{\omega}\big\{Q_eA(\omega,\textbf{k}) \nn
\\&& \times \Pi_e (\omega,\textbf{k})
Q_eA(\omega,\textbf{k})\big\}. \label{eq:5}
\end{eqnarray}

Subsequently, the action, $\mathcal{S}$ can be averaged to arrive
at

\begin{eqnarray}
\mathcal{S} = \frac{T}{2}\int
d\textbf{k}\sum_{ij}\big\{A_i(\omega,\textbf{k})
\Pi_{ij}A_j(\omega,\textbf{k})\big\}. \label{eq:6}
\end{eqnarray}

The averaging was carried out by utilizing the Gaussian
integral~\cite{ryder6}, $\int\exp\big[-\big((1/2)(xWx) + Mx +
N\big)\big]d^nx
=\big((2\pi)^{n/2}/\sqrt{det~W}\big)\exp\big[(1/2)MW^{-1}M - N
\big]$. Therefore, the response function is given by

\begin{eqnarray}
\Pi && = \frac{\big(\gamma +
\beta\big)Q_e^2\Pi_s\Pi_h\Pi_e}{\gamma\Pi_s\Pi_h + \beta\Pi_e\Pi_s
+ \beta\Pi_h\Pi_e} \nonumber \\&& - \frac{\big(\gamma +
\beta\big)Q_e^2\Pi_s\Pi_h\Pi_e}{\gamma\Pi_s\Pi_h + \beta\Pi_e\Pi_s
+ \beta\Pi_h\Pi_e} \nonumber \\&& + Q_s^2\Pi_s + Q_h^2\Pi_h -
(Q_s\Pi_s + Q_h\Pi_h)^2(\Pi_s + \Pi_h)^{-1} \nonumber \\&& =
\frac{Q_e^2\Pi_s\Pi_h}{\Pi_s + \Pi_h} \times \nonumber \\&&
\left[\frac{\big(\gamma + \beta\big)\Pi_e(\Pi_h + \Pi_s) -
\big(\gamma + \beta\big)\Pi_e(\Pi_h + \Pi_s)}{\beta\Pi_e(\Pi_h +
\Pi_s) + \gamma\Pi_h\Pi_s}\right] \nonumber \\&& +
\frac{Q_e^2\Pi_s\Pi_h}{\Pi_s + \Pi_h} \nonumber \\&& =
\frac{Q_e^2\Pi_s\Pi_h}{\Pi_s + \Pi_h} \times \nonumber
\\&&  \left[\frac{\big(\gamma + \beta\big)\Pi_e(\Pi_h + \Pi_s) -
\big(\gamma + \beta\big)\Pi_e(\Pi_h + \Pi_s)}{\beta\Pi_e(\Pi_h +
\Pi_s) + \gamma\Pi_h\Pi_s} + 1\right]. \nonumber
\\ \label{eq:7}
\end{eqnarray}

Firstly, if only spinons and holons exist in $ab$-planes where all
$e$ $\rightarrow$ $s$ + $h$, then Eq.~(\ref{eq:7}) directly gives
$\Pi = Q_e^2[\Pi_s^{-1} + \Pi_h^{-1}]^{-1}$. Simply put, the last
two terms, in Eq.~(\ref{eq:4}) which represent, $L_{e
\rightleftharpoons s + h}$ equals 0, which in turn accentuates the
pure CSS phenomenon. If only electrons exist in $ab$-planes, then
$\Pi_h\Pi_s/(\Pi_h + \Pi_s)$ can be substituted with $\Pi_e$ that
eventually gives, $\Pi$ = $Q_e^2\Pi_e$. Note that the above
rearrangement of Eq.~(\ref{eq:7}) using $\Pi_e$ =
$\Pi_h\Pi_s/(\Pi_h + \Pi_s)$ are {\it solely} to show that
Eq.~(\ref{eq:7}) is as it should be and does not violate the $e
\rightleftharpoons s + h$ processes. In other words, the number of
spinons and holons can only be increased with reduction in
electron's number and $\Pi_e$ = $\Pi_h\Pi_s/(\Pi_h + \Pi_s)$ has
been employed {\it a priori}. Apart from the pure spinon-holon and
pure electron phenomena, if one allows the coexistence of
electrons with spinons and holons, then Eq.~(\ref{eq:7}) can be
reduced as

\begin{eqnarray}
\Pi &&= \frac{Q_e^2\Pi_s\Pi_h}{\Pi_s + \Pi_h}\times\nonumber
\\&&\left[\frac{\big(\gamma + \beta\big)\Pi_e(\Pi_s + \Pi_h) -
\big(\gamma + \beta\big)\Pi_e(\Pi_s + \Pi_h)}{\beta\Pi_e(\Pi_s +
\Pi_h) + \gamma\Pi_s\Pi_h} + 1 \right] \nonumber
\\&& = \frac{Q_e^2\Pi_s\Pi_h}{\Pi_s +
\Pi_h}\times\nn\\&&\bigg[\frac{\big(\gamma + \beta\big)\Pi_e(\Pi_s
+ \Pi_h) - \big(\gamma + \beta\big)\Pi_e(\Pi_s +
\Pi_h)}{\beta\Pi_e(\Pi_s + \Pi_h) + \gamma\Pi_s\Pi_h} \nn\\&& +
\frac{\beta\Pi_e(\Pi_s +
\Pi_h)+\gamma\big\{\Pi_s\Pi_h\big\}}{\beta\Pi_e(\Pi_s + \Pi_h) +
\gamma\Pi_s\Pi_h} \bigg] \nonumber
\\&& = \frac{Q_e^2\Pi_s\Pi_h}{\Pi_s +
\Pi_h}\left[\frac{\big(\gamma + \beta\big)\Pi_e(\Pi_s +
\Pi_h)}{\beta\Pi_e\Pi_s + \beta\Pi_e \Pi_h +
\gamma\Pi_s\Pi_h}\right] \nonumber
\\&& = \frac{\big(\gamma + \beta\big)Q_e^2\Pi_s\Pi_h\Pi_e(\Pi_s + \Pi_h)}{(\Pi_s +
\Pi_h)(\beta\Pi_e\Pi_s + \beta\Pi_e \Pi_h + \gamma\Pi_s\Pi_h)}
\nonumber
\\&& = \frac{\big(\gamma + \beta\big)Q_e^2\Pi_s\Pi_h\Pi_e}{\beta\Pi_e\Pi_s + \beta\Pi_e \Pi_h +
\gamma\Pi_s\Pi_h}. \nonumber \\&& = \big(\gamma +
\beta\big)Q_e^2[\beta\Pi_s^{-1} + \beta\Pi_h^{-1} +
\gamma\Pi_e^{-1}]^{-1}.\label{eq:8}
\end{eqnarray}

Notice that $\Pi_s\Pi_h$, indicated with $\{...\}$ in the fifth
line has been substituted with $\Pi_e(\Pi_s + \Pi_h)$ that
satisfies Ioffe-Larkin formula. This substitution means some of
the electrons ($\Pi_e$) are converted to spinons ($\Pi_s$) and
holons ($\Pi_h$) or vice versa, so as to allow the coexistence
among electrons, spinons and holons (F-CSS liquid). After applying
the linear-response theory, one can arrive at

\begin{eqnarray}
\sigma^{-1} &=& \beta\big[\sigma_s^{-1} + \sigma_h^{-1}\big] +
\gamma\sigma_e^{-1}. \label{eq:9}
\end{eqnarray}

\begin{eqnarray}
\sigma^{-1}_{ab} &=& \sigma_s^{-1} + \sigma_h^{-1} +
\gamma\sigma_e^{-1}. \label{eq:10}
\end{eqnarray}

\begin{eqnarray}
\sigma^{-1}_{c} &=& \beta\big[\sigma_s^{-1} + \sigma_h^{-1}\big] +
\sigma_e^{-1}. \label{eq:11}
\end{eqnarray}

Equations~(\ref{eq:10}) and~(\ref{eq:11}) are precisely in the
form of Eq.~(\ref{eq:1}), because $\beta$ = 1 in $ab$-planes
whereas $\gamma$ = 1 in $c$-axis. Importantly, in $ab$-planes,
$\gamma$ $<$ 1 and $\beta$ = 1 whereas in $c$-axis, $\beta$ $<$ 1
and $\gamma$ = 1. On the contrary, in the pure 2D CSS region with
invalid $s$ + $h$ $\rightleftharpoons$ $e$ processes, $\gamma$ = 0
and $\beta$ = 1 in $ab$-planes while $\gamma$ = 1 and $\beta$ = 0
in $c$-axis. Meaning, the spinons and holons that are confined in
the $ab$-planes are literally independent of the electrons in
$c$-axis, which automatically satisfies the original Ioffe-Larkin
action given in Eq.~(\ref{eq:2}). On the other hand, averaging the
$S_{ind}$ will lead one to the expressions

\begin{eqnarray}
\Pi = \frac{Q_e^2\Pi_s\Pi_h}{\Pi_s + \Pi_h} +
Q_e^2\Pi_e.\label{eq:12}
\end{eqnarray}

\begin{eqnarray}
\sigma^{-1} = \frac{1}{\big[\sigma_s^{-1} +
\sigma_h^{-1}\big]^{-1} + \sigma_e}. \label{eq:13}
\end{eqnarray}

Equation~(\ref{eq:13}) implies that electrons flow is independent
of spinons and holons.

\subsection{3. Analysis}

One can take the suitable limits, as given below in order to
analyze the differences between Eqs.~(\ref{eq:9})
and~(\ref{eq:13}) respectively.

\begin{eqnarray}
\lim_{\sigma^{-1}_e \rightarrow \infty} \sigma^{-1} = \infty,~
\lim_{\sigma^{-1}_e \rightarrow 0} \sigma^{-1} = \sigma_s^{-1} +
\sigma_h^{-1}. \label{eq:14}
\end{eqnarray}

\begin{eqnarray}
\lim_{\sigma^{-1}_e \rightarrow \infty}\sigma^{-1} = \sigma_s^{-1}
+ \sigma_h^{-1},~ \lim_{\sigma^{-1}_e \rightarrow 0}\sigma^{-1} =
0. \label{eq:15}
\end{eqnarray}

Note that the stated Eqs.~(\ref{eq:14}) and~(\ref{eq:15}) are
specifically for underdoped superconducting cuprates, however, the
term $\sigma^{-1}_e$ corresponds to the ionization energy based
Fermi-Dirac statistics (iFDS). Equation~(\ref{eq:14}) suggests
that all components (electrons, spinons and holons) must
superconduct so as to give a 3D superconductivity. Whereas, the
limits in Eq.~(\ref{eq:15}) point out that superconductivity can
be achieved if any of the two phases (electron or spinon-holon)
superconducts. The latter equation also implies that pure CSS is
independently stable in 2D system, opposing the instability due to
additional kinetic energy (KE) scenario calculated by
Sarker~\cite{sarker7}. Add to that, Varma {\it et
al}.~\cite{varma9,batlogg9b} have also discussed that unlike in
1D, the conductivity of pure CSS phase in 2D is rather
irreversible without additional KE. As for the overdoped cuprates,
one can describe the transport properties namely, resistivity,
Hall resistance and Lorenz ratio without employing the CSS
mechanism.~\cite{andrew10,andrew11,andrew12}. Basically, iFDS
derived in the
Refs.~\cite{andrew10,andrew11,andrew12,andrew13,andrew14} has been
employed for the latter work. Apart from that, iFDS is also found
to be viable to determine the electronic properties of
Ba-Sr-Ca-TiO$_3$ ferroelectrics~\cite{andrew14},
ferromagnets~\cite{andrew15} and Carbon nanotubes~\cite{andrew16}.

In conclusion, two possible conductivity rules in the 2D
superconducting systems have been discussed. Coexistence among
spinons, holons and electrons requires their respective
resistivities in series. In certain underdoped high-$T_c$
cuprates, Eq.~(\ref{eq:9}) is more appealing physically than
Eq.~(\ref{eq:13}).

\section*{Acknowledgments}

The author is grateful to Arulsamy Innasimuthu, Sebastiammal
Innasimuthu, Arokia Das Anthony and Cecily Arokiam of CMG-A for
their financial assistances and hospitality.

\end{document}